# A Fault Tolerant Mechanism for Partitioning and Offloading Framework in Pervasive Environments

Nevin Vunka Jungum[1], Nawaz Mohamudally[1] and Nimal Nissanke[2]

[1] School of Innovative Technologies and Engineering, University of Technology Mauritius
La Tour Koenig, Pointe-aux-Sables, Mauritius

[2] School of Computing, Information Systems and Mathematics, London South Bank University
London, UK

**Abstract**
Application partitioning and code offloading are being researched extensively during the past few years. Several frameworks for code offloading have been proposed. However, fewer works attempted to address issues occurred with its implementation in pervasive environments such as frequent network disconnection due to high mobility of users. Thus, in this paper, we proposed a fault tolerant algorithm that helps in consolidating the efficiency and robustness of application partitioning and offloading frameworks. To permit the usage of different fault tolerant policies such as replication and checkpointing, the devices are grouped into high and low reliability clusters. Experimental results shown that the fault tolerant algorithm can easily adapt to different execution conditions while incurring minimum overhead.

## 1. Introduction

A mobile pervasive environment consists of users interacting with mobile devices connected to stationary devices, desktops, servers or other mobile devices wirelessly. Due to mobility of users, frequent network disconnections have become a normal characteristic, and as a consequence, this results in failure of any mobile distributed system running in such environment affecting negatively the reliability of the latter. Several fault tolerance mechanisms [1][2] have been proposed to solve the reliability problem in distributed computing systems. Almost all proposed techniques consider environments with wired homogeneous computational devices. Thus, they are difficult to adapt in a wireless heterogenous mobile computing environment as opposed to grid computing systems. Fault tolerance is a process of reinstating the normal or an acceptable behavior of a system. In pervasive computing environments, network disconnection in the middle of the execution of a task is frequent. It is due to, mainly, mobility of users. For example, if user A's device act as a participating device in a cluster of devices collaborating to execute an application partitioned and offloaded from user B's device and the former moves away from the cluster, then a failure is generated.

Hence, this paper proposes a fault tolerant algorithm, using reactive fault tolerant methods, which is an independent component that can be added to any offloading framework. The fault tolerant component takes as input the different tasks offloading schedules from the existing offloading systems [3][4][5] and ensure the complete application execution by the application of different fault tolerant policies.

The rest of the paper is structured as follows: Section 2 discusses some existing works in the area of fault tolerance in distributed computing systems; Section 3 presents the modeling of the application, device and reliability models; Section 4 describes in detail the calculation of reliability level of devices, presents an algorithm to cluster devices based on reliability level, discusses the replication and checkpointing policies and describes a possible implementation. The fault tolerant algorithm is also presented and discussed along with an analysis of the time complexity of the latter. Section 5 describes the experimental setup, the design and implementation of a simulator and the evaluation and analysis of the results; and finally, Section 6 summarizes the paper.

## 2. Related Works

Two reactive fault tolerance mechanisms often used are checkpointing and replication. Using checkpointing, snapshot of an application state is taken at a pre-define time interval and the latter is saved on disk. The system reliability is determined by the time elapsed between two checkpoints. However, in the case of replication, no snapshot of application state is saved. Actually, a replication of the application is executed in parallel on other computational devices to ensure complete processing of task.

The paradigm of distributed computing encompasses grid computing, mobile grids, cluster computing among many others. In such systems, the computational resources loosely coupled are connected by means of a network that requires the management of fault tolerance to ensure the system stability, robustness and reliability. The authors in [6] made a performance comparison between dynamic load balancing (DLB) and job replication (JR) on distributed systems robust level. A measure statistic Y and a corresponding threshold value Y* were provided, such that DLB consistently outperformed JR, and the reverse is true while Y<Y*. In [7] the authors proposed an incremental checkpoint and restart model for high performance computing (HPC). So as to minimize the overhead of checkpointing, the method performs a set of incremental checkpoints between two full checkpoints by only saving the address space that has changed since the last checkpoint. However, the fault tolerance mechanism in distributed computing systems only take into consideration failures of devices as the wired networks are stable. But when considering mobile pervasive environments whereby computational devices are mostly mobile and networks are mostly wireless, then existing mechanism fall short. A replication-based algorithm [8] that uses the Weibull distribution for mobility analysis was proposed to approximate the number of replicas so as to maintain a level of fault tolerance for mobile grid systems. A onefold fault tolerance policy is applied by most existing algorithms which is not reliable for resource-limited mobile devices collaborating in smart mobile spaces as there is no assurance for the availability of computational devices.

In [9] the authors proposed the k-out-of-n reliability control technique with the objective to achieve energy efficiency and so as to maintain system reliability level for processing and storage of data in mobile cloud. Three factors are considered to estimate failure probability. They are amount of battery life left, mobility of device and application computation requirement. A fault tolerant algorithm was proposed [10] for management of resources in mobile cloud. Multiple fault tolerance methods such as checkpoint and replication were used for multiple clusters of devices depending upon their availability and mobility. But the clustering algorithm takes into consideration only device mobility and rate of utilization. However, more criteria can be considered. In [11] the author categorized devices in mobile grids into groups having identical hardware properties. Tasks sent to low reliability groups; the fault tolerance strategy used is task replication while tasks sent to high reliability groups are offloaded to other device upon failures.

## 3. Modeling of the System

The application is modeled as a directed acyclic graph (DAG) $A = (S, E)$ as shown in Figure 1, where $S$ represents the set of tasks $S_i$ that constitutes the application and $E$ denotes the tasks' dependencies. The execution time of each task is represented by the weight of each node pointed towards it. For example, the tasks $S_1$ and $S_4$ in Figure 1 below takes 8 time units and 6 time units respectively.

| Tasks in S | Task dependencies in E |
|---|---|
| 1 | none |
| 2 | none |
| 3 | 1, 2 |
| 4 | 1, 3, 5 |
| 5 | 2 |
| 6 | 4, 5 |

Table 1 Application task dependency table

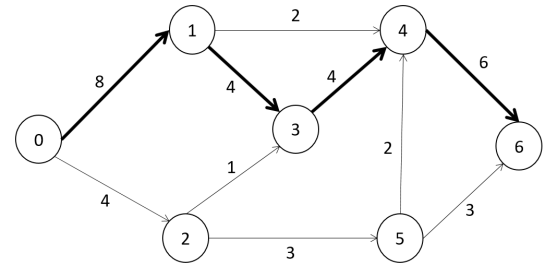

Figure 1 An application modeled as a DAG. Critical path is shown in bold

It is assumed that a mobile task cannot be divided into sub-tasks. Each task has a time variable $T_{tf}^{S_i}$ referred to as the total float. It is the time a task can be delayed without delaying the execution of the overall application. It is computed as the difference between the latest finish time and the earliest finish time. The critical path of a DAG is a path of tasks that has 0 total float, that is, the longest task execution time, hence, the application earliest finish time [12]. Thus, tasks found along the critical path should be subject to the application fault tolerant policies to make sure the application computes on time.

Let suppose a mobile pervasive environment network $M$ consists of $n$ heterogenous computational devices such as smartphones, laptops, desktops, servers, IOT devices and so on. $m_i \in M(i = 1, 2, \ldots, n)$ designates the $i^{th}$ device. The hardware specifications of a device are modeled as follows:
- the processing speed of a device is denoted as $\mu_i$,
- the rate of utilization of the processor is represented by $Q_i$.
- $I_i^{wifi}$ is a binary indicator indicating whether the device is equipped with a WIFI communication medium. Later, this can be further be extended with $I_i^{cell}$ and $I_i^{bt}$ to represent cellular and Bluetooth communication interfaces respectively.

For example, if $I_i^{wifi} = 1$ and $I_i^{ether} = 0$, this means that device $m_i$ has a WIFI interface but no ethernet port for communication.

- $B_i^{wifi}$ and $B_i^{ether}$ represents the bandwidth of the communication medium on device $m_i$. The speed of each communication medium is computed as the product of the binary indicators $I_i$ and bandwidth and the addition of current network latency.
- the earliest available time of a device $m_i$ is $T_i^{avail}$ which reflects the actual load of the latter.
- $T_i^r$ represents the time duration between two failures of device $m_i$. The mobility of the devices in the environment permits the collection of $T_i^r$ of each device.

Reliability of a device implies its availability. Just as a side note, this paper does not focus on the mobility and trajectory of the devices. To get a device $m_i$ available time, the mean time between failures (MTBF) of the latter is considered as $T_i^r$.

The MTBF of a device can be computed using the Weibull distribution. It is a versatile distribution since it can take on the features of other types of distributions (such as normal, exponential and so on) by adjusting its shape parameter, $\beta$. A 2-parameter Weibull distribution probability density function is as follows:

$$f(t) = \frac{\beta}{\eta}\left(\frac{t}{\eta}\right)^{\beta-1} e^{-\left(\frac{t}{\eta}\right)^{\beta}} \quad (1)$$

$\beta$ is the shape parameter and $\eta$ is the scale parameter estimated using historic data from the device connection time to the network and its duration.

## 4. Proposed Fault Tolerant Algorithm

The fault tolerant mechanism along with other components to enable the offloading process is depicted in Figure 2.

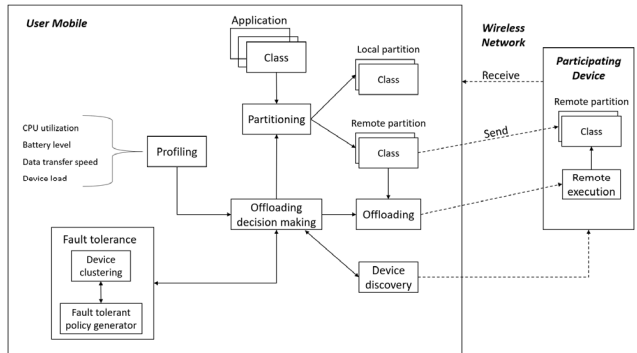

Figure 2 The fault tolerant mechanism along with other components of the offloading framework

The interaction diagram of the fault tolerant component along with the code offloading engine, source device and participating devices is shown in Figure 3.

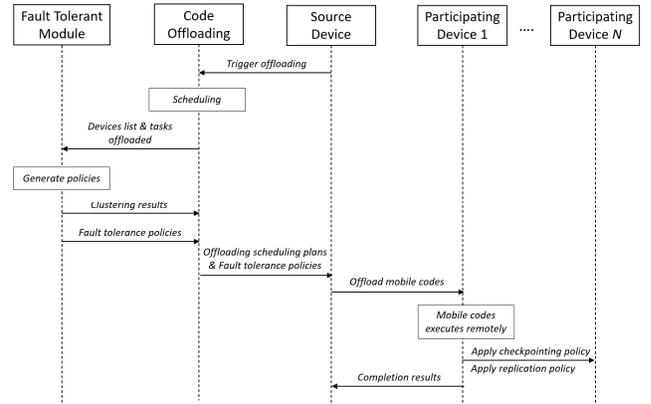

Figure 3 The interaction diagram of proposed fault tolerant mechanism

The offloading decision making component output a list of tasks to be offloaded and their corresponding host devices. We refer to this output as the task offloading scheduling plans (TOSP). The device clustering module clusters the participating devices and the policy generator binds the relevant fault tolerance policy to each task after decoding the TOSP. Three criteria are considered. They are the performance of the device, its availability and its data transfer speed within the network.

### 4.1. Clustering Devices into High and Low Reliability Clusters

The three criteria used to determine the reliability of devices where either replication or checkpointing fault tolerant policy will be applied are the computing capability of the device, its availability and its data throughput in the network. The computing capability of a device $m_i$ is computed as:

$$C_i = \delta_i * \varphi_i \quad (2)$$

where $C_i$ is the computing capability of the latter, $\delta_i$ is the CPU speed and $\varphi_i$ is the CPU utilization rate. The mobility and battery lifetime of the device are considered to determine its availability. The availability of the device is the amount of time it is connected to the network. The longer it is connected to the network and the higher its battery life, this implies, the higher its availability. Hence, it is obtained as follows:

$$A_i = YT_i^r * Z\psi_i \quad (3)$$

where $T_i^r$ represents the device available time from the mobility model and $\psi_i$ represents the percentage of battery remaining. $Y$ and $Z$ are weight factors that can be adjusted depending the context and system requirements and $Y + Z = 1$.

Both fault tolerant policies, replication and checkpointing, needs fast data transfer with minimum energy consumed. Thus, communication with other devices in the environment is taken into consideration. The communication capacity of the device is calculated as follows:

$$\mu_i = B_i - (d * v) \quad (4)$$

where $B_i$ represents the total bandwidth of the device, $d$ represents the data transfer rate and $v$ represents the number of existing connections to other devices we assume a fixed data transfer rate for all connections. Thus, $\mu_i$ reflects the capacity remaining to service future devices.

The three criteria defined, that is, computation capability, availability and communication capacity of the device, are all combined to determine the reliability degree of the latter. The technique of combining all three criteria helps in reducing the computational complexity specially in the event of a very large number of devices. As such, the reliability $R_i$ of a device $m_i$ is calculated as follows:

$$R_i = \alpha_{cpu} C_i * \alpha_{batt} A_i * \alpha_{conn} \mu_i \quad (5)$$

where $\alpha_{cpu}$, $\alpha_{batt}$ and $\alpha_{conn}$ are weight factors and $\alpha_{cpu} + \alpha_{batt} + \alpha_{conn} = 1$.

All devices that potentially may involve in the fault tolerance process will have to be categorized in high reliability and low reliability clusters. Because of the heterogeneity of the devices in the environment, a onefold fault tolerant strategy might not be suitable for all conditions. For example, consider a high reliability cluster consisting of high reliable devices, by applying the replication policy to such cluster will result in more operating overhead than checkpointing. On the other hand, consider a low reliability cluster consisting of low reliable devices, applying the checkpointing policy may create unnecessary storage of snapshots.

Several clustering algorithms exist to group similar entities in one cluster and dissimilar ones to another one. We are using the k-means clustering [13] which is a prototype-based clustering technique that tries to identify a certain $k$ number of clusters specified by the user.

**Procedure**

The first step is the selection of $k$ initial centroids, where $k$ is a parameter specified by the user, that is, the number of clusters desired. Each reliability value $R_i$ of device $m_i$ is assigned to the closest centroid and each group of reliability values $R_i$ assigned to a centroid is called a cluster. Based on the reliability value $R_i$ assigned to the cluster, the centroid of each cluster is updated.

**Algorithm 1** Reliability Clustering Algorithm
1. Choose the number of reliability clusters ($k = 2$) and get the $R_i$ values
2. Place the centroids $c_1, \ldots, c_k$ randomly
3. For each reliability value $R_i$ :
   - find the nearest centroid ($c_1, \ldots, c_k$)
   - assign $R_i$ to that cluster
4. For each cluster $j = 1, \ldots, k$
   - new centroid = mean of all $R_i$ assigned to that cluster
5. Repeat steps 4 and 5 until convergence or no further change in the assignment of $R_i$ data points
6. Return low and high reliability clusters
7. End

### 4.2. Replication and Checkpointing Policies

The task replication is offloaded to the same cluster as the selected participating device, originally scheduled to process the offloaded task as per the code offloading module, in order to maintain the offloading benefits in terms of execution time and energy consumption [3][18] and load balancing [5]. Having unnecessarily a large number of replications may result in an inefficient redundancy of resources along with a high energy overhead. To resolve this issue, a score is calculated for each device based on the execution time for the replication, the failure rate and the number of directly connected devices.

$$score(m) = YT^{S_i} * Z\left(\frac{S_i^{fail}}{S_i^{total}}\right) * \lambda \frac{d^{conn}}{d^{total}} \quad (6)$$

where $Y$, $Z$ and $\lambda$ are adjustable weight factors and $Y + Z + \lambda = 1$. $T^{S_i}$ denotes the task completion time. And $S_i^{fail}$ and $S_i^{total}$ represents the number of tasks failed and the total number of tasks executed by the device. $d^{conn}$ and $d^{total}$ represents the number of devices that $m$ is connected and the total number of devices in the cluster respectively. Then, the device with the lowest score is, in this configuration, the most reliable one in the cluster and is chosen to offload the replica of the task.

Applying the checkpointing policy implies saving a snapshot of the task being executed periodically. Determining the frequency of checkpoints is important as it creates additional network traffic for the wireless communications, thus, more energy usage and checkpointing also incurs time overhead. Therefore, the frequency needs to be calculated based on the time taken by the checkpointing operation and failure rate. The technique proposed by the author in [16] is used to compute the checkpointing frequency as follows:

$$T_c = \sqrt{2T_s T_f} \quad (7)$$

where $T_c$ is the time between two checkpoints, $T_s$ is the checkpointing time and $T_f$ is the time between failures.

## 4.3. The Fault Tolerant Algorithm

The proposed fault tolerant algorithm is listed in Algorithm 2 below.

---

**Algorithm 2** Proposed Fault Tolerant Algorithm

---

**FaultTolerantProcess** $(M, S, S_{offload})$

**for all** $(m \in M)$ **do**
    Calculate $R_i$
**for all** $(m \in M)$ **do**
    $\{M_{HR}, M_{LR}\} \leftarrow (M, R_i)$

$\{S_{cp}\} \leftarrow critical\_path\ (S)$

**for all** $(s \in S_{offload})$ **do**
    **if** $(s \notin S_{cp})$ **then**
        *offload task s to schedule device*
    **elseif** $(s \in S_{cp})$ **then**
        **if** $(s(m) \in M_{LR})$ **then**
            *calculate score of devices in* $M_{LR}$
            $\sim m \leftarrow$ *lowest score device*
            $\sim m \leftarrow S_{replica}$
        **elseif** $(S(m) \in M_{HR})$ **then**
            $T_c \leftarrow$ *calc checkpoint frequency*
            $s \leftarrow$ *append checkpoint with* $T_c$

---

$M$ represents the list of devices in the environment, $S$ denotes the list of all tasks and $S_{offload}$ represents the list of tasks to be offloaded. All these three variables are taken as input by the fault tolerant algorithm. First the reliability value $R_i$ of each device $m$ is calculated. All devices $M$ are then categorized into low reliability $M_{LR}$ and high reliability $M_{HR}$ clusters. $S_{cp}$ is the set of tasks found in the critical path. All tasks that need to be offloaded are checked if they are in the critical path $S_{cp}$. Any task not found in the critical path are offloaded to its scheduled device for execution. For each task found in the critical path, their corresponding offloading scheduled device $m$ is check if the latter is in the low reliability or high reliability cluster. For the low reliability cluster, the score of all devices in that cluster is calculated and the device $\sim m$ with the lowest score is assigned the task replica $S_{replica}$. For the high reliability cluster, the checkpoint is calculated $T_c$ and appended to the task $s$.

Suppose there are $m$ number of devices in the environment and $n$ number of tasks are processed by the fault tolerant algorithm. The first step consists of calculating the reliability $R_i$ of each device in $M$ thus taking $O(m)$. The reliability clustering algorithm takes $O(m * k * I * d)$, where $k$ is the number of clusters, $I$ is the number of iterations until convergence and $d$ is the number of attributes. Since each device has only one attribute which is $R_i$ and the number of clusters is 2, that is, low reliability and high reliability clusters, therefore, the time complexity can be reduced to $O(m * I)$ and hence to $O(m)$. The scoring calculation for the replication policy costs $O(m)$ while the calculation of the frequency of checkpointing takes $O(1)$. Thus, the maximum cost for using the fault tolerant policy for $n$ tasks is $O(mn)$. Therefore, the algorithm overall time complexity is given as $O(m + mn)$.

## 5. Evaluation

### 5.1. The Setup

A number of simulations are performed so as to evaluate the performance of the proposed fault tolerant algorithm. Three metrics are considered for this evaluation and they are the application completion time, the fault tolerant algorithm overhead cost and the number of control messages for fault tolerance. Conventional fault tolerant algorithms are compared with the results obtained to better situate the performance of the proposed algorithm.

A simulator is implemented based on the INET Framework [14] to simulate the devices in the mobile network. INET is an open-source model suite for wired, wireless and mobile networks running on top of OMNeT++ [15] which is an event discrete simulator. Different topologies with varying bandwidth and delay values for each communication link are generated. Table 2 and 3 describes the task parameters list and device parameters list for the simulation respectively.

| Parameters | Value |
|---|---|
| Number of applications | 50 |
| Length of computation (no of instructions) | 20000 – 100000 |
| Size of data (MB) | 0.5 – 10 |

Table 2 Task parameter values for the simulation

| Parameters | Value |
|---|---|
| CPU Speed (MIPS) | 1000 – 100000 |
| Total time available (s) | 1000 – 30000 |
| Weibull (shape & scale) (failure point) | $\alpha = 1.21, \beta = 94.08$ |
| Bandwidth WIFI (MBps) | 0.9 – 1.2 |
| Number of devices | 20 – 50 |

Table 3 Device parameter values for the simulation

We designed and implemented a $\pi$-calculator to estimate to a certain extent the value of $\pi$. The workloads used in this simulation consists of a set of applications represented by different randomly generated DAGs. A task of an application is represented by a vertex in a DAG. Each task has to calculate the value of $\pi$ and takes as input the number of times to run the approximations and the number of decimal places desired.

Each vertex has also two values associated with it, that is, the amount of computation and data size. The two values are generated from within a specified range. The device parameters listed in Table 3 shows that the latter's processing speed is measured in Million Instructions Per Second (MIPS). With every tick of the clock, the CPU fetches and executes one instruction. The clock speed is measured in cycles per second, and one cycle per second is known as 1 hertz. This means that a CPU with a clock speed of 2 gigahertz (GHz) can perform two billion cycles per second.

To simulate failures of devices, a 2-parameter Weibull distribution is considered to generate random time between failures and the failure time points are computed by the device start time to the time generated from the distribution. And for realistic mobility patterns, we used the CRAWDAD trace dataset [17] to get the shape and scale parameters of the Weibull distribution.

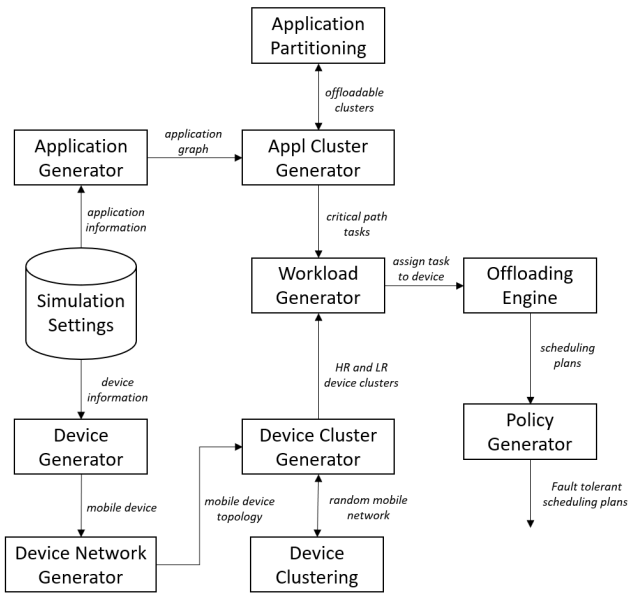

Figure 4 The structural design of the simulator

The simulator, see Figure 4, consists of several modules. Simulation configurations are stored in the simulation settings database or file. And the latter is accessed by the application generator and device generator to obtain simulation data. Random application directed acyclic graphs (DAGs) are generated by the application generator that simulates the mobile tasks and their dependencies. Using the device information from the simulation settings module, the device generator output data for mobile devices. And the device cluster generator uses the former along with the network generator to generated the high and low reliability clusters.

The workload generator gets the tasks on the critical path and the list of high and low reliability device clusters and assign each task to a device. The offloading engine here implements the existing mobile offloading algorithms and send the schedule plans to the policy generator. The latter applies the relevant fault tolerant policies to the offloading schedule plans.

## 5.2. Results and Analysis

The weight factors $Y = 0.2$, $Z = 0.6$ and $\lambda = 0.2$ are used for the experiments. We evaluated the performance of the proposed fault tolerant algorithm with 500 randomly generated applications DAGs. Each application graph is generated with random number of edges and vertices. Each task is bind with a data size and the amount of computation. The output is then compared to other basic fault tolerant strategies such as checkpointing only policy $C^*$, replication only policy $R^*$ and lastly with a no-fault tolerant policy $NoFT^*$, that is, no fault tolerance policy is applied to the scheduling plans. Two performance experiments are considered. Experiment 1 analyze the effect of device availability reflecting on its median time between failures (MTBF). Experiment 2 assesses the consequence of task computation on the fault tolerant algorithm's performance.

**Experiment 1**

In this experiment, the fault tolerant algorithms performance is evaluated. The MTBF is used to denote the availability of the device. The Weibull distribution is used to generate the failure time for each device. Figure 5 shows the average application completion time. We can see that the proposed fault tolerant algorithm outperformed the other three strategies.

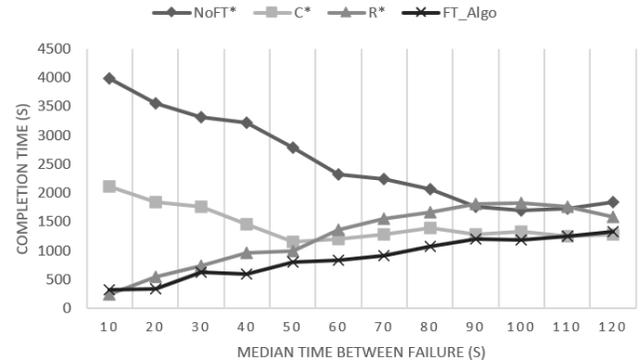

Figure 5 Application completion time based on different device availability

Notice when the MTBF is small (for example, 10 and 20), it implies that the availability of devices is very high and the $NoFT^*$ strategy results in the worst completion time of 4000 seconds compared to other. Whereas $R^*$ and the proposed fault tolerant algorithm ($FT\_Algo$) strategies have amongst the lowest completion time. This is because when the failure occurs more often, the other two strategies, that is, $C^*$ and $NoFT^*$ keeps restarting the task execution and that results in a higher overall application completion time than $R^*$ and $FT\_Algo$. Also, when the MTBF is low, the $FT\_Algo$ applies the replication policy thus the result is similar to the $R^*$ strategy. When the MTBF is above 90, all strategies seem to generate stable results. It is because the devices are more and more reliable, the $R^*$ strategy generates more redundancy overhead for transmitting the replica than $C^*$ as depicted in Figure 6.

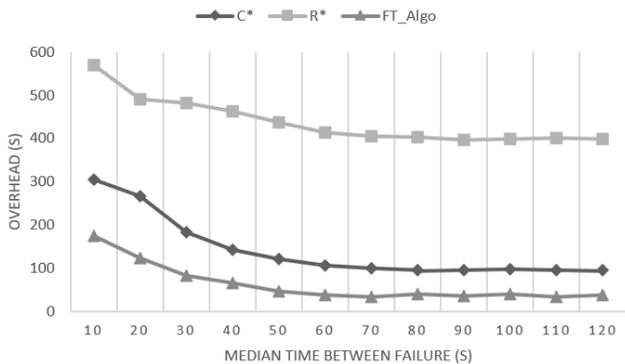

Figure 6 Overhead cost based on different device availability

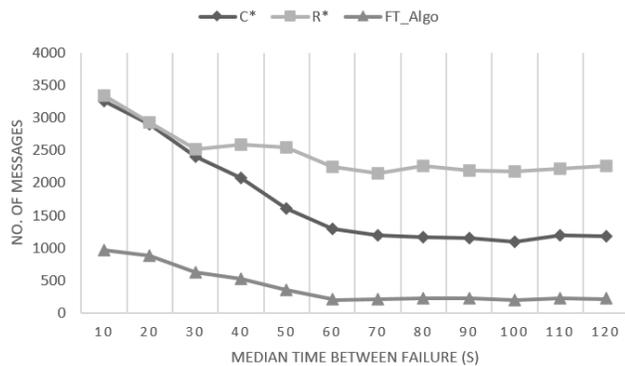

Figure 7 Number of messages based on different device availability

As the MTBF increases, this implies, more time is available for the tasks to be completed before failures occur. Hence, a decrease in the overhead for all strategies. As illustrated in Figures 6 and 7, the overhead and number of messages generated by the overhead tends to decrease and stabilize, which is in line with the overall application completion time.

**Experiment 2**

In this second experiment, the algorithms are evaluated with DAGs having varying amount of task computation requirements in terms of MIPS, to analyze the fault tolerant algorithm performance. MTBF from 90s to 120s are used for the devices.

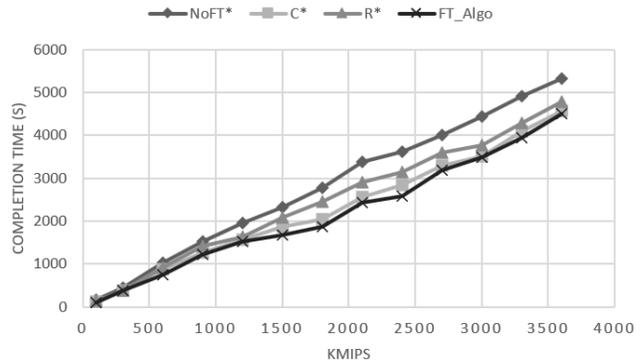

Figure 8 Application completion time based on different task computation requirements

Figure 8 shows the application completion time for the four strategies. The application completion time for all strategies is similar when the amount of computation is low. Even when the computation size grows, all four strategies application completion time grows similarly. The fault tolerant algorithm outperforms all other three strategies and $NoFT^*$ records the worst performance.

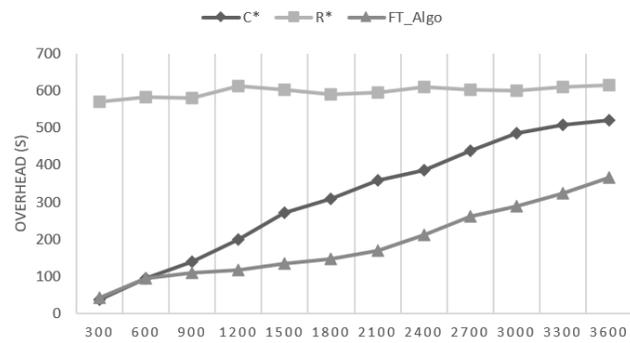

Figure 9 Overhead cost based on different task computation requirements

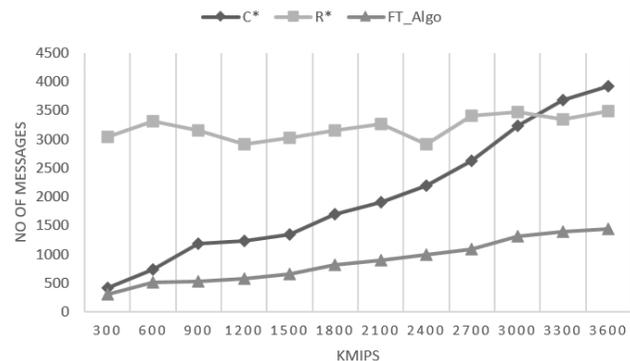

Figure 10 Number of messages based on different task computation requirements

Figures 9 and 10 shows the overhead and number of control messages incurred by the $FT\_Algo$, $R^*$ and $C^*$ algorithms. As the $R^*$ strategy only generates replicas for all tasks, the overhead and number of messages did not fluctuate much. On the contrary, as the task execution gets longer, more checkpointing operations are performed, thus resulting in an increase in overhead and control messages for the *FT_Algo* and the checkpointing strategies.

The results from the different experiments demonstrate that the proposed fault tolerant algorithm can adapt, its policies, to different execution conditions, such as different device availability and different task computation requirement.

## 6. Summary

The mobility of users makes the mobile partitioning and offloading system susceptible to failures. Since most existing fault tolerant algorithms for distributed systems concentrates mainly on device crash failures, their adaptability to pervasive environment that consists of frequent wireless network failures seem difficult and inadequate. Thus, in this paper, we presented a fault tolerant algorithm that helps in consolidating the efficiency and robustness of the partitioning and offloading frameworks. To permit the usage of different fault tolerant policies such as replication and checkpointing, the devices are grouped into high and low reliability clusters. Experiments result shown that the fault tolerant algorithm can easily adapt to different execution conditions while incurring minimum overhead.

## References

[1] Gil Neiger and Sam Toueg, "Automatically increasing the fault-tolerance of distributed systems", in proceedings of the seventh annual ACM Symposium on Principles of distributed computing, pp 248–262, 1988.

[2] N. Xiong, Y. Yang, M. Cao, J. He and L. Shu, "A Survey on Fault-Tolerance in Distributed Network Systems," 2009 International Conference on Computational Science and Engineering, Vancouver, BC, pp. 1065-1070, 2009.

[3] Nevin Vunka Jungum, Nawaz Mohamudally and Nimal Nissanke, "Application Partitioning for Offloading in Mobile Pervasive Environments", in proceedings of The 10th International Conference on Emerging Ubiquitous Systems and Pervasive Networks, EUSPN 2019, November 4-7, 2019, Coimbra, Portugal.

[4] Nevin Vunka Jungum, Nawaz Mohamudally and Nimal Nissanke, "Device Selection Decision Making using Multi-Criteria for Offloading Application Mobile Codes", in proceedings of The 6th IEEE International Conference on Advanced Computing and Communication Systems, ICACCS 2020, March 6-7, 2020, TamilNadu, India.

[5] Nevin Vunka Jungum, Nawaz Mohamudally and Nimal Nissanke, "A Dynamic Load Balancing Algorithm for Distributing Mobile Codes in Multi-Applications and Multi-Hosts Environment", in proceedings of the International Journal of Computer Science Issues, Vol. 17, Issue 4, July 2020.

[6] M. Dobber, R. van der Mei, and G. Koole, "Dynamic load balancing and job replication in a global-scale grid environment: A comparison", IEEE Transactions on Parallel and Distributed Systems, vol. 20, no. 2, pp. 207–218, February 2009.

[7] N. Naksinehaboon, Y. Liu, C. Leangsuksun, R. Nassar, M. Paun, and S. L. Scott, "Reliability-aware approach: An incremental checkpoint/restart model in hpc environments", in Proceedings of 2008 Eighth IEEE International Symposium on Cluster Computing and the Grid, May 2008, pp. 783–788.

[8] A. Litke, D. Skoutas, K. Tserpes, and T. Varvarigou, "Efficient task replication and management for adaptive fault tolerance in mobile grid environments," Future Generation Computer Systems, vol. 23, no. 2, pp. 163 – 178, 2007.

[9] C. Chen, W. Bao, X. Zhu, H. Ji, W. Xiao, and J. Wu, "Agile: A terminal energy efficient scheduling method in mobile cloud computing," Transactions on Emerging Telecommunications Technologies, vol. 26, no. 12, pp. 1323–1336, 2015.

[10] J. Park, H. Yu, H. Kim, and E. Lee, "Dynamic group-based fault tolerance technique for reliable resource management in mobile cloud computing," Concurrency and Computation: Practice and Experience, vol. 28, no. 10, pp. 2756–2769, 2016.

[11] S. Choi, M. Baik, J. Gil, S. Jung, and C. Hwang, "Adaptive group scheduling mechanism using mobile agents in peer-to-peer grid computing environment," Applied Intelligence, vol. 25, no. 2, pp. 199–221, 2006.

[12] J. James E. Kelley, "Critical-path planning and scheduling: Mathematical basis," Operations Research, vol. 9, no. 3, pp. 296–320, 1961.

[13] J. MacQueen et al., "Some methods for classification and analysis of multivariate observations," in Proceedings of the fifth Berkeley symposium on mathematical statistics and probability, vol. 1, no. 14. Oakland, CA, USA., 1967, pp. 281–297.

[14] "OMNeT++ Discrete Event Simulator", retrieved from omnetpp.org

[15] "INET Framework", retrieved from inet.omnetpp.org

[16] J. W. Young, "A first order approximation to the optimum checkpoint interval," ACM Communications, vol. 17, no. 9, pp. 530–531, Sep. 1974.

[17] D. Kotz, T. Henderson, I. Abyzov, and J. Yeo, "CRAWDAD dataset dartmouth/campus (v.2009-09-09)," retrieved from crawdad.org/dartmouth/campus/20090909/syslog, traceset: syslog.

[18] Nevin Vunka Jungum, Nawaz Mohamudally and Nimal Nissanke, "Partitioning Application using Graph Theory for Mobile Devices in Pervasive Computing Environments", in proceedings of The 13th International Conference on Mobile Systems and Pervasive Computing, MobiSPC 2016, August 15-18, 2016, Montreal, Quebec, Canada.

## Acknowledgments

We are grateful to our colleague Richard Wilson {rwilson73@student.mtsac.edu} for his contribution in conducting the simulation of the proposed algorithm using the INET Framework simulation software.